# Green Steganalyzer: A Green Learning Approach to Image Steganalysis

Yao Zhu[1], Xinyu Wang[1], Hong-Shuo Chen[1], Ronald Salloum[2] and C.-C. Jay Kuo[1*]

[1] *University of Southern California, Los Angeles, CA, USA*
[2] *California State University, San Bernardino, CA, USA*

ABSTRACT

A novel learning solution to image steganalysis based on the green learning paradigm, called Green Steganalyzer (GS), is proposed in this work. GS consists of three modules: 1) pixel-based anomaly prediction, 2) embedding location detection, and 3) decision fusion for image-level detection. In the first module, GS decomposes an image into patches, adopts Saab transforms for feature extraction, and conducts self-supervised learning to predict an anomaly score of their center pixel. In the second module, GS analyzes the anomaly scores of a pixel and its neighborhood to find pixels of higher embedding probabilities. In the third module, GS focuses on pixels of higher embedding probabilities and fuses their anomaly scores to make final image-level classification. Compared with state-of-the-art deep-learning models, GS achieves comparable detection performance against S-UNIWARD, WOW and HILL steganography schemes with significantly lower computational complexity and a smaller model size, making it attractive for mobile/edge applications. Furthermore, GS is mathematically transparent because of its modular design.

*Corresponding author: Yao Zhu, yaozhu@usc.edu

*Keywords:* Image Steganalysis, Green Learning, Saab Transform, Decision Fusion.

## 1  Introduction

Image steganography aims to embed hidden information in certain source images called cover images. Secret information is concealed in either the pixel domain by slightly modifying the pixel value or by tweaking the DCT coefficients in JPEG-coded images, resulting in stego images. The most secure steganographic technique today is content-adaptive steganography. With this approach, modification of pixels is likely to happen in complex regions such as edges and texture regions, thus making it more difficult to differentiate cover and stego images.

Image steganalysis is a technique to differentiate stego and cover images. The development of more secure steganography algorithms has led to the need for more powerful steganalysis methods. Before the advent of deep learning (DL), steganalysis was developed by using a feed-forward design. Traditional steganalysis usually consist of three major steps: 1) noise residual computation, 2) feature extraction, and 3) binary classification. In the first step, a variety of heuristic high-pass filters are used for noise residual computation, aiming to suppress image content. In the second step, based on noise residuals, higher order statistics are analyzed and selected as features. In the third step, a machine-learning-based classifier (or an ensemble of several of them) is trained to make final decision.

One famous traditional steganalysis method operating in the spatial domain is the Spatial Rich Model (SRM) [8]. It builds a rich model by taking the union of diverse sub-models learned from quantized patches of noise residuals. SRM was shown to be effective against several simpler steganographic schemes under different payloads. However, SRM does not perform well against more secure steganography schemes such as content-adaptive steganography. Furthermore, SRM has an extremely large feature number, leading to a high computational cost.

Due to the great success of DL in computer vision, DL-based steganalysis has received attention since 2015, e.g., Qian-Net [18], Xu-Net [22], Ye-Net [24], Yedroudj-Net [25], Zhu-Net [29], DRN [21], and GBRAS-Net [19]. Although low-, mid-, and high-frequency components all play a role in computer vision tasks such as object classification and detection, the high-frequency information is more relevant to the detection of weak stego-signals in image steganalysis. Thus, DL-based methods follow the first step in traditional steganalysis and use a high-pass filter (HPF) for image preprocessing. HPF can be viewed as a fixed-parameter layer. It converts raw images into noise residual images as the desired input to deep neural networks (DNNs). Through a careful architectural design, the model parameters of a DNN can be automatically determined by the backpropagation algorithm. End-to-end optimized DNNs yield better detection performance than traditional steganalysis methods. However, their underlying decision mechanism is mathematically obsecure. Furthermore, they have a large number of trainable parameters (i.e., a large model size) and demand higher computational complexity in both training and inference.

The shortcomings of traditional and DL-based steganalysis methods motivate this work. Here, we propose an energy-efficient and mathematically transparent steganalysis method and call it the Green Steganalyzer (GS). GS adopts a modular design based on the green learning paradigm [6, 10, 11, 12, 13]. It is worthwhile to emphasize that we abandon end-to-end optimization as done in DL-based methods but go with the modular design for the purpose of interpretability.

GS consists of three modules: 1) pixel-based anomaly prediction, 2) embedding location detection, and 3) decision fusion for image-level classification. In the first module, GS decomposes an image into patches, adopts Saab transforms [13] for feature extraction from patches, and conducts self-supervised learning to predict an anomaly score of the patch center. In the second module, GS analyzes the anomaly scores of a pixel and its neighborhood to find pixels of higher embedding likelihood. In the third module, GS focuses on pixels of higher embedding probabilities and fuses their abnomaly scores to make final image-level decision.

Compared with traditional and DL-based steganalysis methods, GS has the following three major advantages.

1. High Detection Performance
   We compare the detection error rates of various steganalysis methods against S-UNIWARD, WOW and HILL steganography schemes under two payloads in Sec. 4.2. It is observed that GS out-
performs SRM (a traditional non-DL-based steganalyzer) by a
significant margin. GS also outperforms many DL-based stegan-
alyzers. Its performance is comparable with two state-of-the-art
DL-based steganalyzers (i.e. Zhu-Net and GBRAS-Net).

2. Small Model Size and Low Complexity
   We compare the model sizes and the floating point operation
   numbers (FLOPs) per pixel in the inference stage of DL-based
   steganalyzers and GS in Sec. 4.3. It is shown that GS has a
   substantially smaller mode size and a much lower computational
   complexity.

3. Mathematical Transparency
   As compared with DL-based steganalyzers, GS is mathematically
   transparent due to its modular design. One can gain a clear
   understanding of each individual module.

The rest of the paper is organized as follows. We give a high-level
overview on three image steganalysis approaches; namely, traditional,
DL-based, and the proposed green-learning-based (GL-based) steganal-
lyers in Section 2. Then, we elaborate on the proposed GS in Section 3.
Experimental results on the detection performance and the model sizes
and complexity analysis are presented in Section 4. Finally, concluding
remarks and future extensions are given in Section 5.

## 2   Review of Related Work

Related previous work is reviewed in this section. First, we conduct a
brief survey on traditional and DL-based image steganalyzers in Sec-
tions 2.1 and 2.2, respectively. Then, we present the emerging green
learning paradigm in Section 2.3.

### 2.1   *Traditional Image Steganalysis*

Traditional steganalysis methods use hand-crafted filters to extract a
set of basic features from images. Afterward, statistical tools, such as the histogram or the co-occurrence matrix, are used to derive more advanced features. Finally, image features are fed into a machine learning classifier for the final decision of stego or cover images.

Fridrich *et al.* [8] proposed the Spatial Rich Model (SRM) by considering diverse relationships between pixels. Rich submodels were constructed from the noise component of images. Features from all submodels were concatenated to form a high-dimensional feature vector and fed into ensemble support vector machine (SVM) classifiers to yield the final decision. SRM has proven to be successful against a wide spectrum of steganographic schemes.

Lu *et al.* [17] used the Fisher criterion to reduce the dimension of steganalytic feature vector. They analyzed the separability of single-dimension and multiple-dimension spatial-domain features by using the Euclidean distance. Furthermore, they used the Fisher criterion to select feature components with best separability as the final steganalytic features. Tang *et al.* [20] assigned different weights to pixels based on their embedding probabilities in the feature extraction process. That is, pixels with higher embedding probabilities were assigned larger weights. Their scheme was effective especially under a low embedding rate (i.e., lower than 0.20 bpp) among traditional image steganalysis methods.

### 2.2 DL-based Image Steganalysis

#### 2.2.1 Earlier Work

State-of-the-art image steganalysis methods are dominated by DL-based solutions. Qian *et al.* [18] proposed one of early neural network models, called GNCNN, for image steganalysis. It contained a learnable convolutional neural network (CNN) model with a fixed high-pass filter as its preprocessing layer. It utilized the Gaussian function as nonlinear activation in the convolutional layers. As compared with classical steganalyzers, GNCNN was the first one to automate the feature-extraction and classification steps in a unified system. It achieved performance comparable with that of classical methods against HUGO, S-UNIWARD and WOW three steganographic schemes under various payloads from 0.3 bpp to 0.5 bpp. Xu *et al.* [22] also used a fixed high-pass filter layer to obtain noise residuals as the input to learnable neural networks. Then, they computed the absolute values of features from 5 groups of convolutional modules and used $Tanh$ as the activation func-tion. Ye *et al.* [24] proposed a solution called the Ye-Net. It utilized
the filter banks from SRM as the initialization weights of the first con-
volutional module. Furthermore, it incorporated the selection-channel
aware (SCA) knowledge in the design of the CNN architecture and
adopted a novel activation, called the Truncated Linear Unit (TLU),
to better suit the nature of stego-noise ($\pm1$). Yedroudj *et al.* [25]
proposed another CNN-based model called Yedroudj-NET. It brought
together the merits of its predecessors; e.g., the use of predefined high-
pass filter banks from SRM as a preprocessing step and the adoption of
both absolute Value (ABS) and TLU activations in the convolutional
module.

### 2.2.2   Recent Work

DNNs have become popular in recent years due to their better per-
formance. Several advanced DL-based image steganalyzers have been
proposed based on DNNs to achieve higher detection perofrmance at
the expense of a higher computational cost.

Wu *et al.* [21] utilized the ResNet architecture for steganalysis,
which they called DRN (deep-residual-learning-based network). Again,
a high-pass filter was used as a preprocessing step. The residual learn-
ing blocks was adopted to preserve features from weak stego signals.
Boroumand *et al.* [3] proposed a deep residual paradigm called the
SRNet that minimized heuristic design elements in the system. It did
not use any predefined high-pass filter as a preprocessing steo. It dis-
abled the pooling step in the first convolutional blocks to prevent the
information loss from weak stego signals. SRNet is currently one of the
most powerful steganalysis methods for high-detection performance.
However, it suffers from a large model size and high computational
complexity. Zhang *et al.* [29] proposed the Zhu-Net that has $3 \times 3$
kernels and separable convolution operations to reduce the number of
trainable parameters. The spatial pyramid pooling is utilized to en-
hance the representation ability of features through multi-level pool-
ing. More recently, Reinel *et al.* [19] proposed the GBRAS-Net that
combines the merits of SRN and Zhu-Net. GBRAS-Net performs well
while maintaining a relatively small model size.

### 2.3 Green Learning

Green learning [12] aims to provide an energy-efficient and mathematically transparent solution to data-driven learning problems. GL-based models are modularized. They are trained in a feed-forward manner without back-propagation. Their training complexity is quite low that it can be carried out solely on CPU. Their model sizes are small and inference complexity is also modest. As a result, it is suitable for mobile/edge computing.

One key module in GL is "unsupervised representation learning". It exploits the underlying statistics of pixels to derive data-driven transforms such as the Saak transform [11] and the Saab transform [13]. Multi-stage Saab transforms can be cascaded to form a PixelHop system [6] for image classification. Distinct from DL-based models that determine filter weights by back-propagation, GL-based methods derive filter weights by analyzing the correlation structure of a local neighborhood centered at a pixel of interest.

Another key module in GL is "supervised feature learning" [23]. It is a feature selection strategy that chooses a more powerful subset of features by analyzing the entropy loss of each single feature dimension. Entropy loss is used to evaluate the discriminant power of each feature for the classification task. A feature subset with low entropy loss can offer the same or even better performance than the whole feature set. It can significantly reduce the feature dimension for the classifier, which in turn reduces the computational cost and the model size.

GL has been successfully applied to various vision tasks, such as image classification [6, 7], image anomaly localization [26], object tracking [30, 31, 32], image synthesis [1, 14, 15, 16, 33], and 3D point cloud classification, segmentation, and registration [9, 27, 28]. It has also been applied to image forensics such as deepfake video detection [4], GAN-generated fake image detection [34], etc.

In this work, we propose a novel GL-based image steganalyzer called GS. GS does not use heuristically designed high-pass filters as a pre-processor. It does not adopt the end-to-end optimized neural network, either. It differentiates stego and cover images based on anomaly detection of image patches. Although the stego signal is weak in patches, we can leverage pixel-based anomaly scores to zoom into fewer patches for further steganalysis. The proposed GS method is light-weight, mathe-matically transparent, and computationally efficient.

## 3   Green Steganalyzer (GS) Method

We present an overview of the proposed GS method and its rationale in Section 3.1. GS consists of three modules. They are detailed in Sections 3.2, 3.3, and 3.4, respectively.

### *3.1   Solution Overview and Rationale*

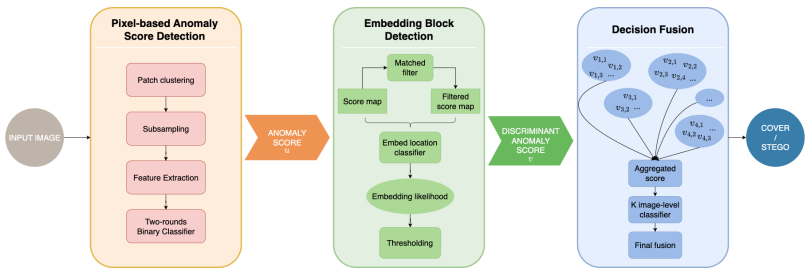

Figure 1: An overview of the proposed GS method.

For traditional steganalysis methods, one of the essential steps is to use predefined (or heuristically designed) filters to extract noise residuals from input images. It helps suppress image content and increase the signal-to-noise ratio (SNR). However, the fixed filter weights and a limited variety of filters cannot capture all of the complex cases in cover images and embedded stego signals. DL-based steganalyzers can learn data-driven filters via backpropagation. They still use predefined high-pass filters in the pre-processing layer. Furthermore, they use a deep network architecture, leading to a large number of trainable parameters and heavy computational complexity.

To address the shortcomings of two prior methodologies, we propose a new pipeline and depict it in Figure 1. A high-level description of each module and its rationale are given below.

- Module 1: Pixel-based Anomaly Prediction
  Motivated by anomaly detection, we consider the stego embedding, especially embedding in smooth regions as 'anomaly'. We aim to predict the 'degree of anomaly' within local regions of images. The 'degree of anomaly' is named as 'anomaly score'. We first decompose an input image into overlapping image patches with stride equal to one. The high patch diversity makes any machine learning task challenging. To mitigate this problem, patches are grouped into multiple sets based on the embedding cost of each steganographic method. Then, patch diversity in a group is reduced. For each group, we train a binary classifier to discern two cases - anomaly patches that contain an embedded stego-signal (labeled by "1") and the corresponding raw patches from cover images (labeled by "0"). We obtain content-adaptive filters through unsupervised representation learning and the filter responses are used as features to the classifer in each group. We conduct an iterative classification scheme to enhance the detection performance, and derive an anomaly score for the central pixel of each patch.

- Module 2: Embedding Location Detection
  We design an embedding location detector to localize potential embedding locations based on anomaly scores. Since the anomaly score of a single pixel is noisy and untrustworthy, it is not reliable to use simple thresholding. We need a better idea. That is, when a pixel is modified by a stegonagraphic scheme, it has an effect not only on its own anomaly scores also on those of its neighboring pixels. These pixels form an "anomaly spot". Then, the center of the anomaly spot is chosen as the embedding location. This strategy finds embedding locations more accurately.

- Module 3: Decision Fusion for Image-Level Classification
  We obtain the anomaly scores of all pixels in Module 1 and embedding pixel locations in Module 2. We aggregate anomaly scores of selected pixels from all groups for ultimate binary image-level decision (a cover or a stego image) in the last module.

We show the data processing pipelines of traditional, DL-based, and the proposed GS steganalysis methods in Figure 2. There differences are summarized below.

1. Both traditional and DL-based steganalyers have a pre-processing step in the beginning of the pipeline. GS does not have this step.

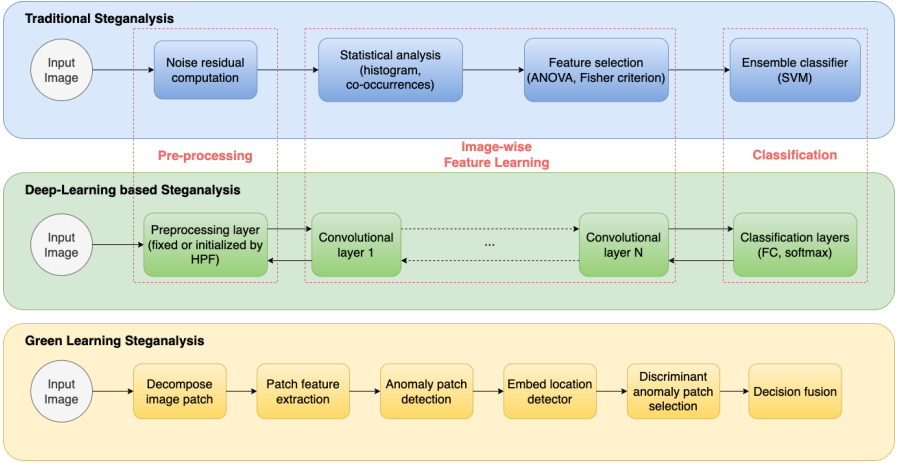

Figure 2: Comparison of the data processing pipelines: (top) traditional image steganalysis, (middle) DL-based image steganalysis, and (bottom) the GS method.

> It derives an anomaly score of each pixel via supervised learning process.

2. Both traditional and DL-based steganalyers examine image-level representations. In contrast, GL works on local regions only (either a patch in Module 1 or a neighborhood region in Module 2). This local processing can be parallelized easily. The memory requirement is much less.

3. For binary image-level classification, traditional steganalyers often leverages the ensemble of multiple classifiers (e.g., the ensemble of SVMs). The ensembled SVM classifier is slow in practice due to the high-dimensional features. DL-based steganalysis methods use the fully-connected (FC) layers and the softmax layers to make final decision. The number of trainable parameters in the FC layers is large. In contrast, GS uses the averaged anomaly score of anomaly spots as features to train multiple lightweight binary classifiers and an ensemble classifier. Its computation cost is much smaller.

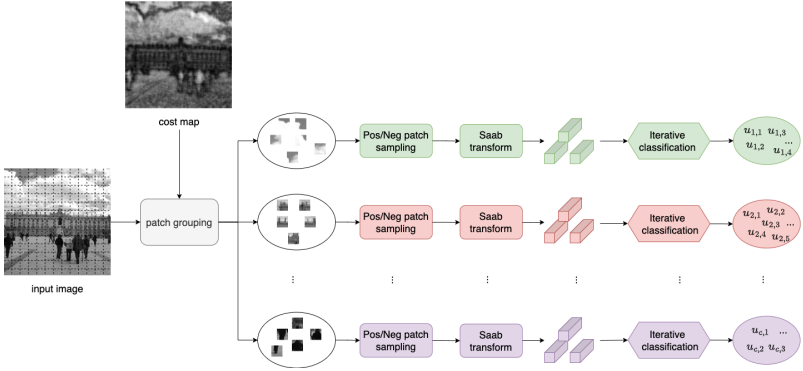

Figure 3: The block diagram of Module 1.

### 3.2 Module 1: Pixel-based Anomaly Prediction

The block diagram of the first module is shown in Figure 3. The goal is to estimate the deviation of a patch of size $P \times P$ from its original one caused by steganographic embedding, where $P$ is a user-selected parameter. The anomaly caused by a stego signal in a smooth region is easier to tell than those in complicated textured regions. To address this challenge, we design a patch anomaly score predictor with the following four steps.

1. **Patch Grouping.** Input images are decomposed into overlapping patches of size $P \times P$ with stride 1. In the experiment, we set $P = 7$. Patches are grouped into multiple sets based on the embedding cost of a steganographic scheme. Each steganographic scheme has a specific way to compute the embedding cost of a pixel. If a pixel has a lower embedding cost, it has a higher embedding probability. Take S-UNIWARD steganography as an example. Its patch-wise cost varies from 0 to 11. We partition patches into groups with narrower cost ranges such as $[0, 1)$, $[1, 2)$, etc. Depending on the steganography algorithm, the group number ranges from 10 to 12.

2. **Positive and Negative Patch Sampling.** For each group, we select positive and negative patch samples in this step. We choose patches from stego images that contain at least one embedding bit and call them positive samples. Then, we find the corresponding patches from cover images and use them as negative samples.

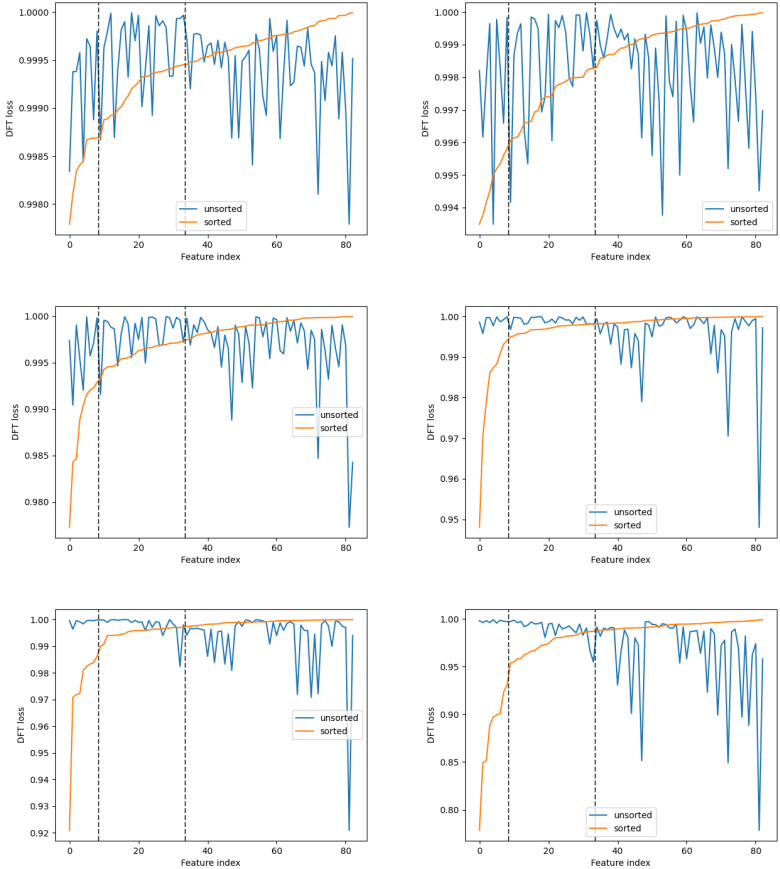

Figure 4: Illustration of the loss curves obtained by the discriminant feature test (DFT) for four exemplar groups, where their embedding costs increase from left-to-right and top-to-bottom. For each subplot, blue and orange curves represent *unsorted and* sorted DFT loss curves, respectively. Two vertical dashed lines in black partition unsorted features into three regions: from $3 \times 3$ filters (left), $5 \times 5$ filters (middle), $7 \times 7$ filters (right), respectively.

3. **Feature Extraction.** We apply three Saab transforms to the patch center [6, 13]. The filter sizes are $3 \times 3$, $5 \times 5$, and $7 \times 7$, respectively. The Saab transform is a variant of the Principal Component Analysis (PCA) transform. It has two types of transform kernels: 1) the DC kernel, which gives the local average of pixels covered by the filter, and 2) AC kernels, which are data-driven kernels obtained by PCA. The reason to have two kernel types is that PCA can only be applied to zero-mean random vectors. By removing the local block mean, the block residual can be treated as a zero-mean random vector so that PCA can be applied. For a Saab transform of size $n \times n$, we have $n \times n$ filters. The filter responses are used as features since they describe the pixel local correlation structure. Compared with filters of smaller sizes, filters of larger sizes are more discriminant in smooth regions but less in complicated regions. The aggregation of filter responses of different filter sizes is beneficial. By concatenating filter responses from three Saab transforms, we have $(3 \times 3) + (5 \times 5) + (7 \times 7) = 83$ features in total.

We select a subset of discriminant features using the discriminant feature test (DFT) [23] to reduce the feature dimension per group. The DFT calculates a loss value for each individual feature dimension. The DFT loss curves of four exemplar groups are shown in Figure 4, where their embedding costs increase from left-to-right and top-to-bottom. In each subplot, the blue and orange curves denote the original and sorted DFT losses, respectively. Two vertical dashed lines in black partition 83 unsorted features into 3 regions: features from $3 \times 3$ filters (the left region), $5 \times 5$ filters (the middle region), $7 \times 7$ filters (the right region), respectively. Generally speaking, features from $7 \times 7$ filters are more discriminant while features from $3 \times 3$ filters are less discriminant. The gap of their discriminability widens as the embedding cost increases. To reduce the parameter number, we select 15 feature dimensions in all 10 groups. Selected feature are used to train a binary classifier to discern positive and negative patch samples collected in Step 2.

4. **Iterative Classification.** In the final step, we would like to obtain the anomaly score for each patch center using an iterative classification idea.

   *i) Round-1 Classification.* We first train a binary XGBoost classifiers [5] using labels and features obtained in Steps 2 and 3, re-

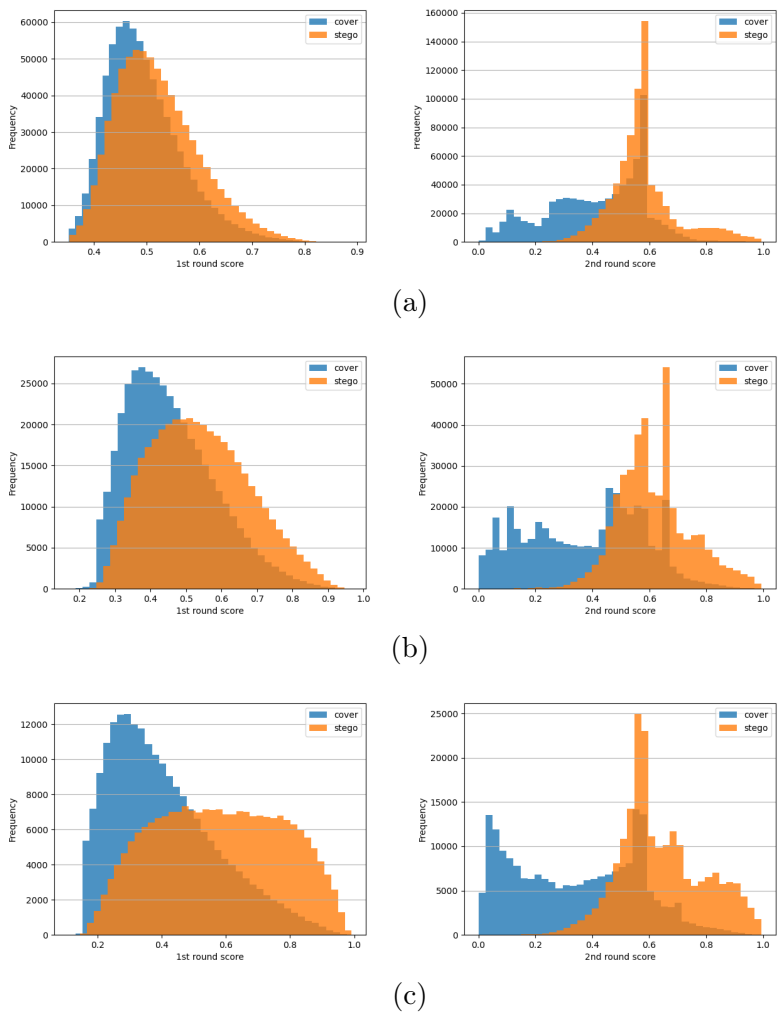

(a)

(b)

(c)

Figure 5: The anomaly score histograms of positive (in orange) and negative (in blue) test samples of three representative groups, where the left column and right column show results of the first- and the second-round XGBoost classifiers.

spectively. In the inference stage, we keep the soft decision whose value lies between $[0, 1]$, which is called the round-1 anomaly score. If the score of a patch is closer to 0 (or 1), it is more likely to be drawn from cover (or stego) images. The predicted anomaly score distributions of positive and negative test samples (drawn from stego and cover images, respectively) are plotted in the left column of Figure 5. The figure has three rows. Each row indiciate a representative group as described in Step 1. It is observed that there is a significant overlap between positive and negative histograms. In words, they cannot be easily separated by one round of classification. To boost the classification performance, we adopt an iterative classification idea using the second-round classification.

*ii) Round-2 Classification.* For each group, we partition positive samples into subgroups accoording to their anomaly scores from Round 1. For example, we can divide their anomaly scores into ten uniform intervals, [0,0.1), [0.1, 0.2), ..., [0.9, 1.0]. Then, positive samples in the same interval form a subgroup. For each subgroup, we gather the corresponding negative samples to build training positive/negative pairs and use them to train the 2nd XGBoost classifier with their groundtruth labels (i.e., 1/0) by following Step 3 and Step 4.i. Then, we merge the predicted anomaly score distributions of positive/negative test samples from the 10 subgroups of the same group and plot them in the right column of Figure 5, where each row compares the positive/negative histograms after Round-1 and Round-2 classifications for three selected representative groups.

We can see the clear advantage of the two-round iterative classification scheme by comparing the left and the right columns in Figure 5. First, the positive and negative histograms are more separated from each other. Second, the highest anomaly scores of stego patches from certain groups with round-1 classification can only reach 0.8 and 0.9 as shown in Figure 5(a) and Figure 5(b), respectively. However, they can be boosted to 1.0 after Round-2 classification.

In the inference stage without image padding, we scan all interior pixels of test images using a window of size $P \times P$ with stride one, extract features using Step 3, and compute the anomaly score for the center pixel using Step 4.

### 3.3    Module 2: Embedding Location Detection

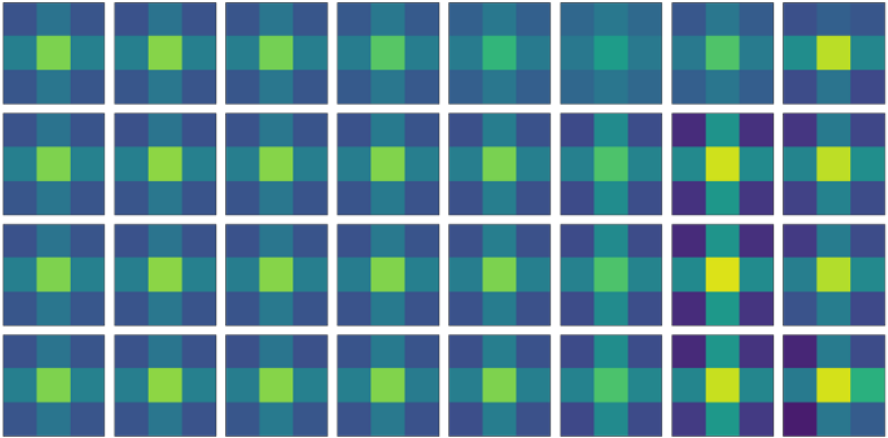

Figure 6: Visualizationof matched filters from 8 different groups (i.e., one column per group) under the S-UNIWARD steganography algorithm with its payload equal to 0.4 bpp (1st row), 0.3 bpp (2nd row), 0.2 bpp (3rd row), 0.1 bpp (4th row). The brighter color indicates a larger value. The embedding cost increases from left to right in the same row (i.e. the same payload).

We get an anomaly score for each pixel and obtain an anomaly score map after Module 1. The purpose of Module 2 is to estimate the embedding location based on the anomaly score map. Its block diagram is shown in Figure 7. It contains the following two major steps.

1. **Anomaly Spot Localization.** As shown in Figure 5, the anomaly score of a single pixel is still noisy, it is not reliable to use simple thresholding to decide the embedding location. Instead, we should consider a set of connected pixels jointly, which form an anomaly spot. Here, we set the size of anomaly spot to $3 \times 3$. We design an anomaly spot localizer for each individual group as follows.

   *i) Positive and Negative Block Sampling.* A block has a size of $3 \times 3$. We collect positive and negative block samples by following the same idea described in Step 2 of Module 1.

   *ii) Matched Filtering for More Discriminant Features.* The anomaly scores of nine pixels in a block of size $3 \times 3$ can be used as features to classify whether it is an anomaly spot through a binary classifier. However, these features are not discriminant enough to ensure good classification performance. To boost the classification performance, we add another set of features using matched filtering. The matched filter is a $3 \times 3$ kernel. We collect the anomaly score maps of all positive samples and conduct element-wise averaging to obtain one matched filter for each group. The coefficient values of representative matched filters are visualized in Figure 6. All of them share similar properties. First, its central coefficient has the largest value, four sides' coefficients are smaller, four corners' coefficients are the smallest. The values of four sides are close to each other, and the values of four corners are also close to each other. This can be explained by the symmetrical property in space. Second, the range of coefficients is wider for a higher embedding cost. The application of matched filtering to the anomaly score map of a block will enhance the difference between positive and negative samples.

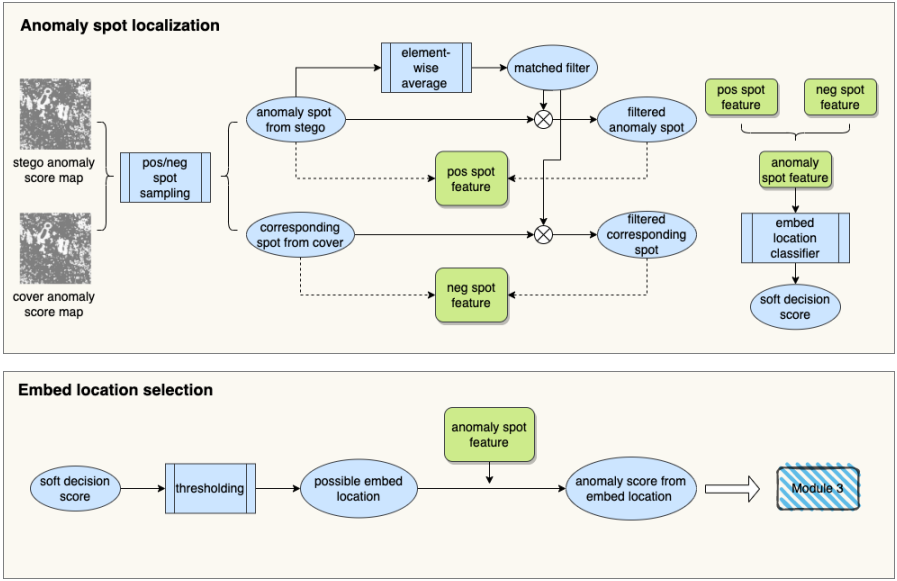

Figure 7: The block diagram of Module 2.

*iii) Classification.* We apply matched filters to anomaly maps to get pixel-wise matched-filter responses for each group. A

block has 9 pixels, and each pixel has its own anomaly score and matched-filter response. Then, we use 9 anomaly scores and 9 matched-filter responses to form an 18-D feature vector, train an XGBoost binary classifier. The soft decision score indicates the likelihood for a block to contain embedding bits.

2. **Embedding Location Selection.** If the soft decision score of a block is higher than a threshold, it serves as a candidate for consideration in the image-level decision. It is called an anomaly spot. The threshold is selected by optimizing the F1 score, which is a measure used to balance false positives and false negatives. The center of the anomaly spot is the detected embedding location.

To show the importance of Module 2, we compare the distributions of anomaly scores from Module 1 and distributions of soft decision scores of anomaly spots from Module 2 for three representative images in Figure 8. First, we want to point out that the number of anomaly spots is significantly less than the number of pixels in test images. Thus, a large number of pixels are already removed in Module 2. Second, we should pay special attention to the right tail of the histogram. Comparing with the distribution Module 1, the distribution in Module 2 is more separable in the right-tail region. The discriminant ability of positive/negative samples after Module 1 is still weak. Yet, they are more distinguishable after Module 2.

### 3.4  *Module 3: Decision Fusion for Image-Level Classification*

Recall that we obtain the anomaly scores for all pixels in Module 1 and select embedding pixel locations in Module 2. In Module 3, we aggregate anomaly scores of selected pixels from all groups. We sort their anomaly scores from the highest to the lowest, and use top $M$ anomaly scores as features for image-level decision. If $M$ is too small, the decision is not reliable. If $M$ is too large, we may include unrelable pixels in the decision. Thus, a proper value of $M$ has to be determined based on the validation dataset. We conduct a grid search for the optimal M values from 100 to 1000, with step size 50. We show the accurate classification rate as a function of $M$ for a representative validation image. We see that the accuracy goes higher as $M$ becomes larger. However, it is not an monotonically increasing curve. In the experiment, we choose five $M$ values, $M_i$, $i = 1, \cdots, 5$ and conduct an XGBoost classifier for each $M$ value. Each XGBoost will give a binary decision - stego or cover. The final image-level decision is made by the majority vote.

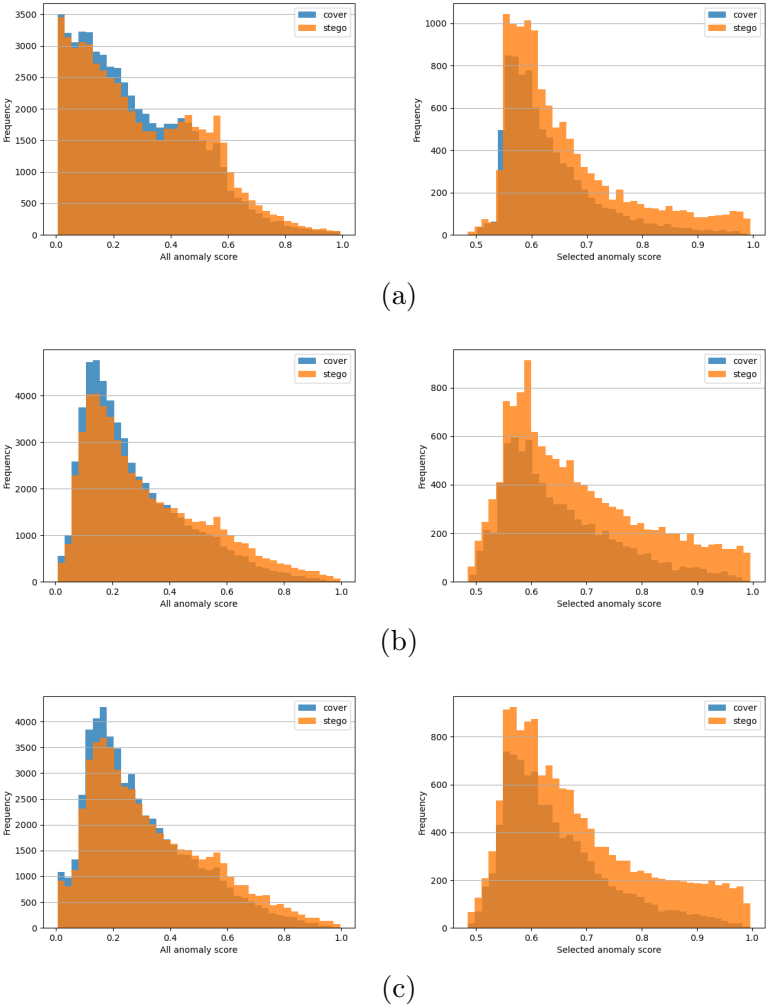

Figure 8: Comparison of distributions of anomaly scores from Module 1 (in the left column) and distributions of soft decison scores of anomaly spots from Module 2 (in the right column) for three representative images (in three rows).

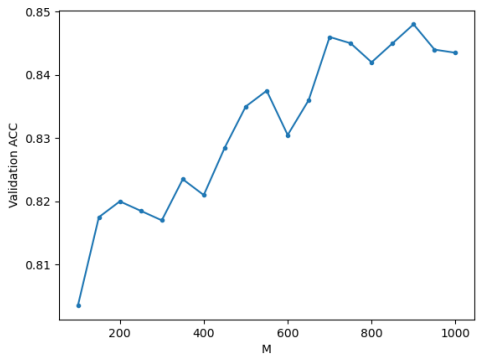

Figure 9: The accurate classification rate as a function of $M$ values applied to the validation dataset.

## 4  Experiments

We conduct experiments to demonstrate the effectiveness and efficiency of the proposed GS scheme in this section. The experimental setup is first described in Sec. 4.1. Then, we compare the effectiveness of GS with 7 benchmarking methods in terms of the detection error rate in Sec. 4.2. Finally, we compare the efficiency of GS with 5 benchmarking methods by examining their model sizes and computational complexity in Sec. 4.3.

### 4.1  Experimental Setup

The settings of our experiments are given below.

   **Dataset.** Experiments are conducted on the BOSSbase v1.01 dataset [2]. It contains 10,000 8-bit gray-scale images of resolution $512 \times 512$. They are stored of the uncompressed Portable Gray Map (pgm) format. These images are acquired with several cameras. They cover diverse natural scenes with various texture characteristics. The BOSSbase dataset has been widely used as a test dataset for digital image staganalysis. We split 10,000 BOSSbase images evenly into 50% training data and 50% testing data. For fair comparison with other benchmarking methods in [18], [22], [24], [29], we resize raw images of resolution $512 \times 512$ to new images of resolution $256 \times 256$ in both training and test datasets and evaluate the GS method on resized images.

**Steganographic Schemes.** We consider S-UNIWARD, WOW and HILL three content-adaptive steganographic schemes and implement them using the Syndrome-Trellis Codes (STC). For each steganography scheme, stego images are generated with two payloads - 0.2bpp and 0.4bpp. In steganography, the less the payload, fewer bits are embedded, making stego images more difficult to detect.

**Benchmarking Methods.** We compare GS with the following representative steganalysis methods:

- Traditional Methods: Spatial Rich Model with Ensemble classifier (SRM+EC) [8];

- Earlier DL-based Methods: QianNet [18], Xu-Net [22], Ye-Net [24], and Yedroudj-Net [25];

- Recent DL-based Methods: Zhu-Net [29] and GBRAS-Net [19].

All of them are tested under the same train/test split as ours.

**Evaluation Metrics.** We use the averaged detection error rate,

$$P_E = \frac{1}{2}(P_{FA} + P_{MD}), \tag{1}$$

as the performance evaluation metric, where $P_{FA}$ and $P_{MD}$ are the false alarm probability and the missed detection probability, respectively. Besides detection accuracy of steganalyzers, we also examine their model sizes measured by the number of model parameters and computational complexity measured by the number of floating-point operations (FLOPs) in the inference stage.

### 4.2 Detection Performance Evaluation

First, we compare detection error rates of GS against 7 benchmarking methods on the S-UNIWARD and WOW datasets with payloads of 0.2 bpp and 0.4 bpp in Table 1. The best and second best results are highlighted in bold and with an underline, respectively. As shown in the table, the two recent DL-based methods (i.e. Zhu-Net and GBRAS-Net) and our GS rank the top three. Among the three, GBRAS-Net has the best performance for both S-UNIWARD and WOW steganographic embeddings. GS achieves the second best for S-UNIWARD and the third best for WOW. Zhu-Net ranks the third for S-UNIWARD and the second for WOW.

Table 1: Comparison of detection error rates ($P_E$) against S-UNIWARD and WOW steganographic schemes at paylods equal to 0.2 bpp and 0.4 bpp, where the best is in bold and the second best is underlined.

| Method | Payload | | | |
|---|---|---|---|---|
| | S-UNIWARD 0.2 bpp | S-UNIWARD 0.4 bpp | WOW 0.2 bpp | WOW 0.4 bpp |
| SRM+EC | 36.6 | 24.7 | 36.5 | 25.5 |
| Qian-Net | 46.3 | 30.9 | 38.6 | 29.3 |
| Xu-Net | 39.1 | 27.3 | 32.5 | 20.7 |
| Ye-Net | 39.9 | 31.3 | 33.1 | 23.3 |
| Yedroudj-Net | 36.5 | 22.6 | 27.7 | 14.9 |
| Zhu-Net | 28.6 | 15.5 | <u>23.1</u> | <u>11.9</u> |
| GBRAS-Net | **26.4** | **12.9** | **19.7** | **10.2** |
| GS (Ours) | <u>27.86</u> | <u>13.63</u> | 24.05 | 12.18 |

Since no error rates are reported by Qian-Net, Xu-Net, Yedroudj-Net and Zhu-Net for the HILL steganography, we only compare GS with Zhu-Net and GBRAS-Net with payloads equal to 0.2 bpp and 0.4 bpp for HILL in Table 2. This is sufficient since GS, Zhu-Net, and GBRAS-Net are the top three performers in Table 1. Again, we observe that GS and Zhu-Net have comparable performance and their performance is inferior to that of GBRAS-Net.

### 4.3 Model Sizes and Computational Complexity

**Model sizes.** We compare the model sizes of GS and three other DL-based steganalyzers in Table 3, where the model size is defined as the number of trainable parameters. The trainable parameters of GS include Saab filter parameters and anomaly patch classifier parameters in Module 1, embedding location classifier parameters in Module 2, and decision fusion classifier parameters in Module 3. For a given steganography scheme, we partition patches and blocks into 10 groups based on its embedding cost. The parameter number of each module is calculated below.

Table 2: Comparison of detection error rates ($P_E$) under the HILL steganography at paylods equal to 0.2 bpp and 0.4 bpp, where the best is in bold and the second best is underlined.

| Method | Payload | |
|:---:|:---:|:---:|
| | HILL 0.2 bpp | HILL 0.4 bpp |
| Zhu-Net | 33.4 | 23.5 |
| GBRAS-Net | **31.5** | **18.1** |
| GS (Ours) | <u>33.13</u> | <u>23.29</u> |

1. Module 1

   There are three parts: Saab filter banks and XGBoost classifiers.

   *a) Saab Filter Banks.* In each group, we train with three Saab filter banks of size $3 \times 3$, $5 \times 5$, and $7 \times 7$. As mentioned in Section 3.2, we select 15 filters among the 83 filters. Among the 15 selected filters, we count the number of filters from $3 \times 3$, $5 \times 5$, $7 \times 7$, respectively, and plot the bar plot in Figure 10. Different groups have different sets of selected filters. They are however consistent for all three experimented steganography algorithms. Based on the statistics in Figure 10, we aggregate the Saab parameters from all groups and have $7,364$ parameters. For different steganography algorithms, we can safely say that our Saab parameters is no more than $8k$. Thus, the number of Saab parameters for all 10 groups is reduced from $(3 \cdot 3 \cdot 9 + 5 \cdot 5 \cdot 25 + 7 \cdot 7 \cdot 49) \cdot 10 = 31,070$ to $8K$ in Module 1.

   *b) XGBoost Classifiers.* For iterative classifiers in Module 1, we use the XGBoost classifier with 100 trees and maximum depth of 2. Each tree has a maximum depth of 2 so that it has at most 10 parameters. There are approximately $1K$ parameters per XGBoost classifier. The first round has 10 classifiers (one for each of the 10 groups). The second round has $10 \times 10 = 100$ classifiers. There are 110 XGBoost classifiers with $110K$ parameters in total.

By combining (a) and (b), the total number of parameters in
Module 1 is equal to $8K + 110K = 118K$.

2. Module 2
   We need 10 XGBoost classifers (one for each of the 10 groups).
   They have the same hyper-parameters as those in Module 1.
   Thus, the number of parameters is $1K \cdot 10 = 10K$.

3. Module 3
   As mentioned in Sec. 3.4, we need 5 classifiers with $M_i$ features,
   $i = 1, \cdots, 5$. Thus, the total number of parameters in Module 3
   is $1K \cdot 5 = 5K$.

Then, the total number of parameters of the GS method is $118K +
10K + 5K = 132K$.

We measure the model sizes of 3 high-performance DL-based ste-
ganalyzers and list them in Table 3. Among them, Yedroudj-Net and
Zhu-Net have larger model sizes because of deep-layer architectures and
denser FC layers. GS has the smallest model size. It is typical to assign
4 bytes to each parameter. Then, the GS model demands 528K byte
memory.

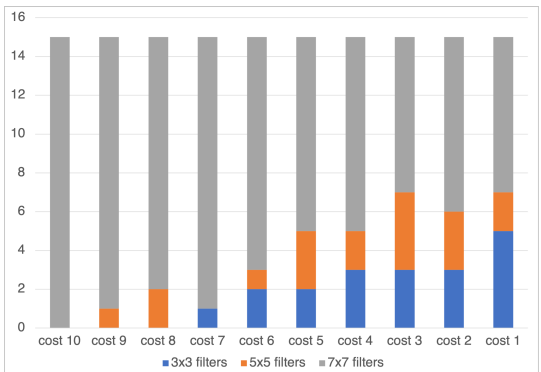

Figure 10: Comparison of selected channel distributions for 10 groups based on
embedding costs, where "cost 1" denotes the lowest embedding cost group, "cost 10"
denotes the highest embedding cost group, and blue, orange, and gray denote filters
of size $3 \times 3$, $5 \times 5$, and $7 \times 7$, respectively.

**Computational Complexity.** We measure the number of floating-
point operations (FLOPs) per pixel in the *inference* stage as an indi-cator of computational complexity. There are several APIs in PyTorch or Keras to measure FLOPs. The numbers of FLOPs for the DL-based methods in Table 3 are measured using the *keras-flops* package for each image and then divided by $256 \cdot 256$, which is the total number of pixels per image.

Table 3: Comparison of model sizes and computational complexities of 6 steganalyzers, where we use "X" to demonstrate the ratio of numbers with respect to the reference (denoted by 1X).

| Method | Number of parameters | KFLOPs/pixel |
|---|---|---|
| Yedroudj-Net | 252,459 (1.91X) | 190.73 (54.03X) |
| Zhu-Net | 277,156 (2.10X) | 45.62 (12.92X) |
| GBRAS-Net | 166,598 (1.26X) | 90.79 (25.72X) |
| GS (Ours) | 132,000 (1.00X) | 3.53 (1.00X) |

Since the GS model is not a neural-network-based model, the *keras-flops* package cannot be directly used. Instead, we compute its number of FLOPs analytically as follows.

1. Module 1

   There are three parts: patch cost calculation, Saab filter banks and XGBoost classifiers. *a) patch cost.* For each pixel location, there needs 8 additions and 1 division, resulting in 9 FLOPs.

   *a) Saab Filter Banks.* For Saab filters of size $n \times n$, the total number of operations per filter is about $2 \times n \times n$ since the inner product of two 9-D vectors involve 9 multiplications and 8 additions. There are three Saab filter sizes. The number of FLOPs for all three Saab filters per pixel is equal to $2 \times (3 \times 3 \times 9 + 5 \times 5 \times 25 + 7 \times 7 \times 49) = 6214$. As explained in both Section 3.2 and Section 4.3 model size part, we select 15 filters among all of them. Based on Figure 10, we calculate number of FLOPs for each group and sum them up to be $2,512$. Thus, the FLOPs/pixel number is actually reduced to $2,512$.

   *b) XGBoost Classifiers.* We conduct Round-1 and Round-2 two XGBoost classifiers at each pixel location. The computational complexity of an XGBoost is a subtraction at each node and one sample will trace only one path. Thus, the complexity for all trees is the tree depth multiplied by the tree number and number of classes. All trees prediction need to be summed up via addition. Thus, the total complexity is equal to $2 \times 100 \times 2 + 100 = 500$.

By combining (a) and (b), the FLOPs/pixel number in Module 1 is $2512 + 500 = 3,012$.

2. Modules 2

   Module 2 computation involves the convolution with matched filters and XGBoost classification. Since we choose anomaly spot size as $3 \times 3$, for each pixel location, convolution needs 9 multiplications and 8 additions. As for XGBoost classifier, the number of FLOPs/pixel is $2 \times 100 \times 2 + 100 = 500$. Thus, the total FLOPs/pixel number in Module 2 is 517.

3. Modules 3

   There are five XGBoost classifiers in Module 3 per image. Thus, the number of FLOPs is equal to $5 \times (2 \times 100 \times 2 + 100) = 25K$. We need to divide this number by $250 \times 250$ pixel locations per image since no image padding is used. The number of FLOPs/pixel is equal to 0.4, which is neglible as compared to the numbers of FLOPs/pixel in Modules 1 and 2.

The total number of FLOPs/pixel of GS is about $3,012 + 517 = 3,529$, which is far less than that of benchmarking DL models.

## 5   Conclusion and Future Work

A GL-based image steganalysis method, called Green Steganalyzer (GS), was proposed in this work. GS is a lightweight modularized image steganalysis method. It contains three modules. First, it assigns an anomaly score to a center pixel of a patch. Next, it studies the relationship of anomaly scores between a pixel and its neighbors to estimate the embedding likelihood of the center pixel. Finally, it selects pixels of higher embedding probabilities and conducts decision that error detection rates of GS are competitive with state-of-the-art DL-based steganalyzers against S-UNIWARD, WOW and HILL three steganographic schemes. At the same time, it demands a smaller model size and lower computational complexity than DL-based methods. Furthermore, GS is mathematically transparent due to its modular design.

As for future extensions, we would like to test GS on the ALASKA 2 dataset. It is a more challenging dataset than BOSSbase v1.01 since it contains more natural scenes and images of various resolutions. DL-based steganlyzers are restrained on a certain input image size because of their architecture design. In contrast, our GS model can handle different image sizes between training and testing dataset. Also, it is desired to improve the detection performance of GS furthermore with slightly higher complexity.